\documentclass[useAMS, usenatbib]{mn2e}

\usepackage{amssymb}
\usepackage{graphicx}
\usepackage{threeparttable}
\usepackage{multirow}
\usepackage{longtable}
\usepackage{subfig}
\usepackage{url}

\newcommand{\mum}{\ifmmode{\rm \mu m}\else{$\mu$m }\fi}

\newcommand{\Msun}{\ensuremath{{\rm M}_{\odot}}}      
    
\newcommand{\chisq}{\ifmmode{\chi^{2} }\else{$\chi^2$}\fi}
\newcommand{\rchisq}{\ifmmode{\chi^{2} }\else{$\chi^2_\nu$}\fi}

\title[Extreme AGB stars in M32]  
{A {\em Spitzer Space Telescope} survey of extreme Asymptotic Giant Branch stars in M32}

\author[O.~C.~Jones et al.]   
{O.~C.~Jones,$^{1,2}$ \thanks{E-mail: ojones@jb.man.ac.uk}
 I.~McDonald,$^{1}$
  R.~M.~Rich,$^{3}$
   F.~Kemper,$^{4}$ 
 M.~L.~Boyer,$^{5,6}$
\newauthor
A.~A.~Zijlstra$^{1}$
and  G.~J.~Bendo$^{1,7}$\\
$^1$ Jodrell Bank Centre for Astrophysics, Alan Turing Building, Oxford Road,  Manchester, M13 9PL, UK.\\
$^{2}$ Space Telescope Science Institute, 3700 San Martin Drive, Baltimore, MD 21218, USA. \\
$^{3}$ Department of Physics \& Astronomy, University of California Los Angeles, Los Angeles, California, USA. \\
$^{4}$ Academia Sinica, Institute of Astronomy and Astrophysics, 11F ASMAB, NTU/AS campus, No. 1 Sec. 4 Roosevelt Road, Taipei 10617, Taiwan.\\
$^{5}$ Observational Cosmology Lab, Code 665, NASA Goddard Space Flight Center, Greenbelt, MD 20771, USA.\\
$^{6}$ Oak Ridge Associated Universities (ORAU), Oak Ridge, TN 37831, USA.\\
$^{7}$ UK ALMA Regional Centre Node.}

\begin{document}

\date{Accepted 2014 October 15.  Received 2014 October 13; in original form 2014 July 21}

\pagerange{\pageref{firstpage}--\pageref{lastpage}} \pubyear{2014}

\maketitle

\label{firstpage}

\begin{abstract}
We investigate the population of cool, evolved stars in the Local Group dwarf elliptical galaxy M32, using Infrared Array Camera observations from the {\em Spitzer Space Telescope}. We construct deep mid-infrared colour-magnitude diagrams for the resolved stellar populations within 3.5 arcmin of M32's centre, and identify those stars that exhibit infrared excess. 
Our data is dominated by a population of luminous, dust-producing stars on the asymptotic giant branch (AGB) and extend to approximately 3 mag below the AGB tip. We detect for the first time a sizeable population of `extreme' AGB stars, highly enshrouded by circumstellar dust and likely completely obscured at optical wavelengths.  The total dust-injection rate from the extreme AGB candidates is measured to be $7.5 \times 10^{-7}$ ${\rm M}_{\odot} \, {\rm yr}^{-1}$, corresponding to a gas mass-loss rate of $1.5 \times 10^{-4}$ ${\rm M}_{\odot} \, {\rm yr}^{-1}$. These extreme stars may be indicative of an extended star-formation epoch between 0.2 and 5 Gyr ago.
\end{abstract}


\section{Introduction} 

Messier 32 (M32) is a dwarf elliptical galaxy in the Local Group and satellite of M31. Our relative proximity to M32 (distance $ 785 \pm 25 $ kpc; \citealt{McConnachie2005}) allows us to detect individual stars. Consequently, this provides us with the nearest opportunity to study the stellar populations of an elliptical galaxy in detail. M32's stellar populations have been extensively studied across many wavelengths (e.g.~\citealt{Baade1944, Rose1985, Grillmair1996, Worthey2004a, Monachesi2011, Sarajedini2012}). However, these populations have not been characterised in the mid-infrared (mid-IR).

It is thought that M32 has two main populations: an intermediate-age, metal-rich  ([Fe/H] = $\sim$+0.1) population thought to be $\sim$2--8 Gyr old \citep{Rose2005, Coelho2009}, and an old (8--10 Gyr) stellar population with slightly sub-solar metallicity  ([Fe/H] = --0.2; \citealt{Monachesi2011}). Additionally, a small number of ancient, metal-poor ([Fe/H] = --1.42) RR Lyrae variables in M32 have recently been identified \citep{Fiorentino2010b,Fiorentino2012}, and a young population of 0.5--2 Gyr stars is probably present at all radii \citep{Monachesi2011}. 
The extremely high surface brightness and sharply peaked brightness profile towards the nucleus of M32 indicates a strong increase in the stellar density. Given the extreme crowding, the stellar populations within the core ($r$ $\lesssim 1'$ of the nucleus) of M32 are incompletely probed. This high surface brightness also prevents studies of the fainter stars in M32.

The unique nature of M32 in the Local Group and the controversies over its origins make M32 a tantalising prospect for stellar population studies (proposed formation scenarios include a true elliptical galaxy or a severely truncated early-type spiral; \citealt{Faber1973,  Bekki2001,  Choi2002, Chilingarian2009, Kormendy2009}). Our immediate goal is to assess the infrared stellar populations of M32 using {\em Spitzer}/IRAC observations and to study the circumstellar dust emission from late-stage AGB stars. 

The Infrared Array Camera (IRAC; \citealt{Fazio2004}) on-board the {\em Spitzer Space Telescope} \citep{Werner2004} provides a unique window to probe M32's cool, luminous stars, whose spectral energy distributions peak in the near-IR and which often have an infrared excess due to thermal emission from circumstellar dust. Indeed, the dustiest, extreme AGB stars are often obscured at optical and near-IR wavelengths, and thus can only be detected by their infrared emission \citep{Meixner2006, Blum2006, Mould2008, Boyer2011, Gerke2013}.


`Extreme' AGB (x-AGB) stars experience a considerable mass loss and are thought to dominate the mass return to the interstellar medium (ISM) in Local Group galaxies \citep{Riebel2012, Boyer2012}. These stars were first identified by \cite{Blum2006} as AGB stars whose near-IR stellar flux is heavily extinguishes by circumstellar dust, thus their K-band magnitude is often fainter than the tip of the RGB \citep[see][]{Boyer2011}.  The criterion \cite{Blum2006} used to select x-AGB star candidates was bases on their location in the [8.0] versus $J-[8.0]$ colour magnitude diagram, where a break is seen in the carbon-star sequence at  $J-[8.0] = 3.5$. Later this selection criteria was modified by \cite{Boyer2011} to focus on stars brighter than the tip of the RGB at 3.6 $\mu$m  with [3.6]--[8.0]$>0.8$; enabling the identification of objects completely obscured at optical and near-IR wavelength. In this paper we adopt the latter selection criteria for x-AGB stars (see Section~\ref{sec:M32CMDs}).

By definition, the x-AGB class represents the high mass-loss rate evolved stellar population in a system. Identifying these sources is important for constraining the dust budget of a galaxy \citep{Riebel2012}. Particularly as the most-extreme sources: AGB stars experiencing the final `superwind', only constitute a small faction of the x-AGB class.

In this paper we describe the results of {\em Spitzer} IRAC and MIPS observations of M32. Our observational data, reduction and photometry is described in Section \ref{sec:M32_Observations}, and we discuss contamination by M31 and foreground stars in Section \ref{sec:starcount}. Infrared colour-magnitude diagrams (CMDs) and luminosity functions are presented in Section \ref{sec:M32results}; we use these to determine the integrated dust return to the ISM (Section~\ref{sec:M32MLRs}). Finally the conclusions are summarised in Section \ref{sec:conclusion}.

\section{Observational data}\label{sec:M32_Observations}

\subsection{Observations} \label{sec:M32obs}

\begin{figure} 
\centering
\includegraphics[trim=2cm 1.5cm 1cm 0cm, clip=true, width=84mm]{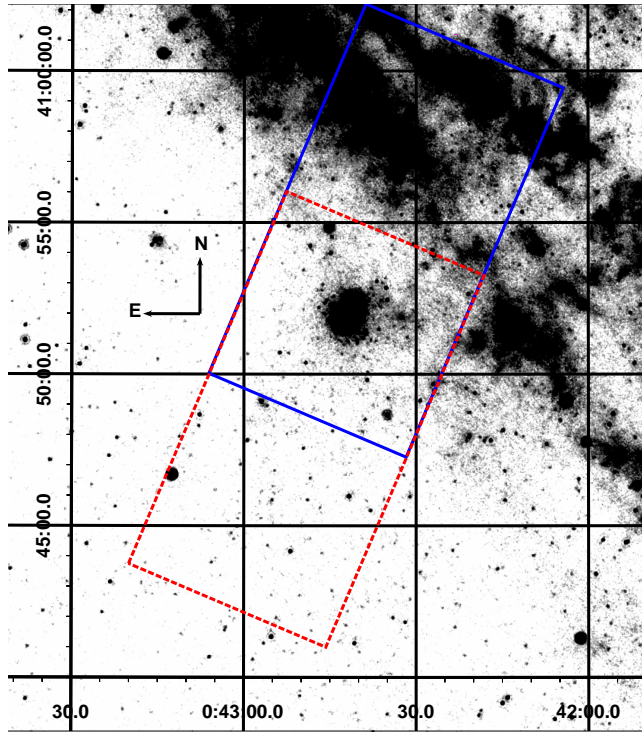}
 \caption[MIPS 24 $\mu$m image of M32 showing the {\em Spitzer} IRAC coverage]{ Location of our {\em Spitzer} IRAC pointings towards M32 superimposed on a MIPS 24 $\mu$m mosaic \citep{Gordon2006, Bendo2012}. The region observed with IRAC [3.6] and [5.8] is shown in red and the IRAC [4.5] and [8] observations are shown in blue.} 
  \label{fig:M32_obs_cov}
\end{figure}

{\em Spitzer} observations of M32 (Program Identification [PID] 3400, PI. M. Rich) were obtained on 18 January 2005 UT with IRAC \citep{Fazio2004} at 3.6, 4.5, 5.8 and 8 $\mu$m.  For each IRAC channel,  92  (5.2'$\times$5.2') frames at 23 dither positions were taken in a cycling pattern with 30-second exposures to build redundancy against outliers and artifacts, resulting in total integration time on M32 of 2760s over most of the map. 
Observations were centred at R.A. $=$ 00$^{h}$42$^{m}$41$^{s}$.6, decl.~$=$ $+$40$^{\circ}$55$^{\prime}$53$^{\prime \prime }$.6 (J2000.0) and cover an area of approximately 6$^{\prime}$$\times$7$^{\prime}$ in all four IRAC channels around the centre of M32, plus an off-field region (offset by one pointing) of the same size to the north-west for IRAC bands 2 and 4  or south-east for IRAC bands 1 and 3. Figure~\ref{fig:M32_obs_cov} shows this field of view overlaid on the MIPS 24-\mum mosaic of M31.
No dedicated background control fields were observed to complement these observations.

\begin{table*}
\caption{Catalogue of Point Sources for M32.}
  \label{tab:M32cat}
  \begin{tabular}{@{}ccccc@{\ }c@{\ }cc@{\ }c@{\ }cc@{\ }c@{\ }cc@{\ }c@{\ }c@{}}
   \hline
   \hline
           &                        & 	     	  &	     & \multicolumn{12}{c}{ Apparent Magnitude} \\   
 SSTM32 \# &  Source ID             & 	   RA	  &  Dec     & \multicolumn{3}{c}{[3.6]}  &  \multicolumn{3}{c}{[4.5]}   & \multicolumn{3}{c}{[5.8]}  & \multicolumn{3}{c}{[8.0]}  \\
   \hline
  0001     &  J004235.04+404839.2   &   10.6460	  & 40.8109  &  13.885  & $\pm$ & 0.051  & 12.694 & $\pm$ & 0.088 & 12.243 & $\pm$   & 0.065  & 11.446 & $\pm$   & 0.063  \\
  0002     &  J004227.59+405300.2   &   10.6150	  & 40.8834  &  15.378  & $\pm$ & 0.241  & 15.043 & $\pm$ & 0.215 & 999.99 & $\pm$   & 999.9  & 15.317 & $\pm$   & 0.125  \\
  0003     &  J004238.90+404830.5   &   10.6621	  & 40.8085  &  15.004  & $\pm$ & 0.184  & 13.739 & $\pm$ & 0.116 & 13.347 & $\pm$   & 0.043  & 12.543 & $\pm$   & 0.055  \\ 
  0004     &  J004241.59+404917.7   &   10.6733	  & 40.8216  &  17.350  & $\pm$ & 0.242  & 16.754 & $\pm$ & 0.208 & 999.99 & $\pm$   & 999.9  & 999.99 & $\pm$   & 999.9  \\
   \ldots  &   \ldots               &    \ldots   & \ldots   & \ldots   &       & \ldots & \ldots &       & \ldots& \ldots &         &\ldots  & \ldots &         & \ldots \\
   \hline
\multicolumn{16}{l}{Notes: The source ID follows the standard Spitzer naming convention, giving the truncated (J2000) coordinates.}\\   
\multicolumn{16}{l}{Only the first entries are shown for guidance, a full version is available in the electronic edition.}
 \end{tabular}
\end{table*}

\subsection{Data Reduction} \label{sec:M32dataReduction}

The raw data was processed by the {\em Spitzer} Science Center (SSC) reduction pipeline version S18.7.0; this removes electronic bias, subtracts a dark image, applies a flat-field correction, and linearises the pixel response. The resulting Basic Calibrated Data (BCD) images were corrected for various instrumental artifacts (e.g.~muxbleeds and column pull-down effects) and combined using the {\sc mopex} reduction package \citep{MakovozMarleau2005} to produce a single mosaic for each channel.

\begin{figure} 
\centering
\includegraphics[width=84mm]{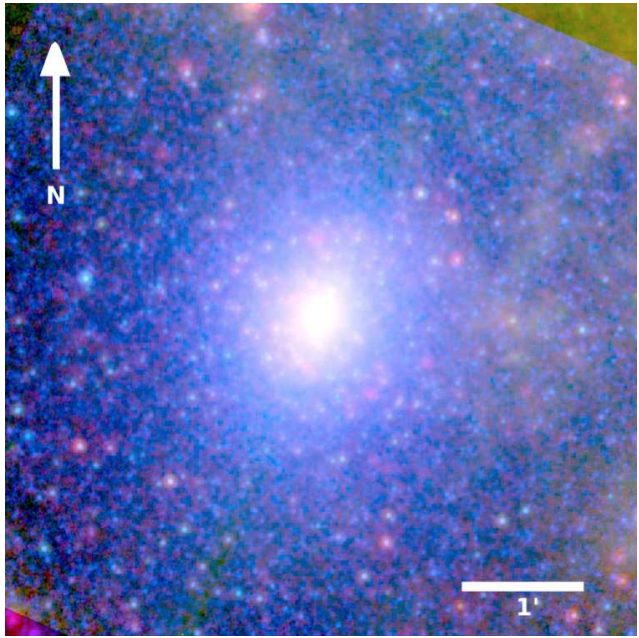}
 \caption[M32 three-colour image]{Three-colour image of M32. Blue is 3.6 $\mu$m, green is 8.0 $\mu$m and red is 24 $\mu$m. Point sources with a dust excess become visible at 8.0 $\mu$m and 24 $\mu$m and appear red in the image.} 
  \label{fig:M32col}
\end{figure}

The {\sc mopex overlap} routine was implemented to match the backgrounds of individual frames in overlapping areas of the images producing a smooth background. The  {\sc mosaicker}, in addition to image interpolation and co-addition, eliminated cosmic rays and other outliers in the data. The final IRAC mosaics are not sub-sampled; thus each image has a pixel scale of 1.22$^{\prime \prime }$ pixel$^{-1}$ and a Point Source Function (PSF) Full With Half Maximum (FWHM) of $\sim$1.9~arcsec.

24~$\mu$m data from the Multiband Imaging Photometer for {\it Spitzer} (MIPS) were used in part of this analysis.  The data, which was originally published by \cite{Gordon2006}, were reprocessed using the MIPS Data  Analysis Tools \citep{Gordon2005} and additional data processing steps described by  \cite{Bendo2012}.  The image has a pixel size of 1.5~arcsec, a PSF FWHM of 6~arcsec, and a calibration uncertainty of 4 per cent \citep{Engelbracht2007}.

Initial inspection of the images revealed a population of clearly resolved point sources at 8 $\mu$m, associated with M32 but not obviously matched at shorter wavelengths.
Figure~\ref{fig:M32col} presents a three-colour image combining the 3.6, 8.0, and 24 $\mu$m observations; the red point sources are clearly visible. Additionally, we see no emission indicative of interstellar dust within M32. This would be particularly evident in the 8.0-$\mu$m map, which is sensitive to non-stellar emission from polycyclic aromatic hydrocarbons (PAHs), or the MIPS 24-$\mu$m mosaic, which is sensitive to cool dust.

\subsection{Photometry}\label{M32photo}

Point sources were extracted from individual IRAC frames and on the mosaicked images using PSF fitting with the {\sc DAOphot ii} and {\sc allstar} photometry packages \citep{Stetson1987}. To determine the shape of the PSF, {\sc DAOphot} requires selection of isolated stars in the field; this is quite challenging in very crowded fields such as M32. PSFs were created from a minimum of 10 isolated bright stars in each IRAC channel, and sources 4$\sigma$ above the local background were chosen for extraction. 

{\sc Allstar} was then used to fit PSFs to all sources found on the image. This is an iterative process that simultaneously fits each source in the frame with the PSF profile and subtracts converged sources from the input image.  As sources are removed, new estimates for the local background are calculated, improving the flux estimates for crowded and faint sources. Sharpness and roundness cuts were employed to eliminate  from the sample extended sources, cosmic rays and unrecognised blends that are broader or narrower than the PSF. Cosmic rays are also eliminated from the point-source list when detections in the same band are combined. The resulting source lists are then cross-correlated with the point-source photometry from the mosaicked data.

\begin{figure*} 
\centering
\includegraphics[trim=0cm 4.22cm 0cm 0cm, clip=true,width=\textwidth]{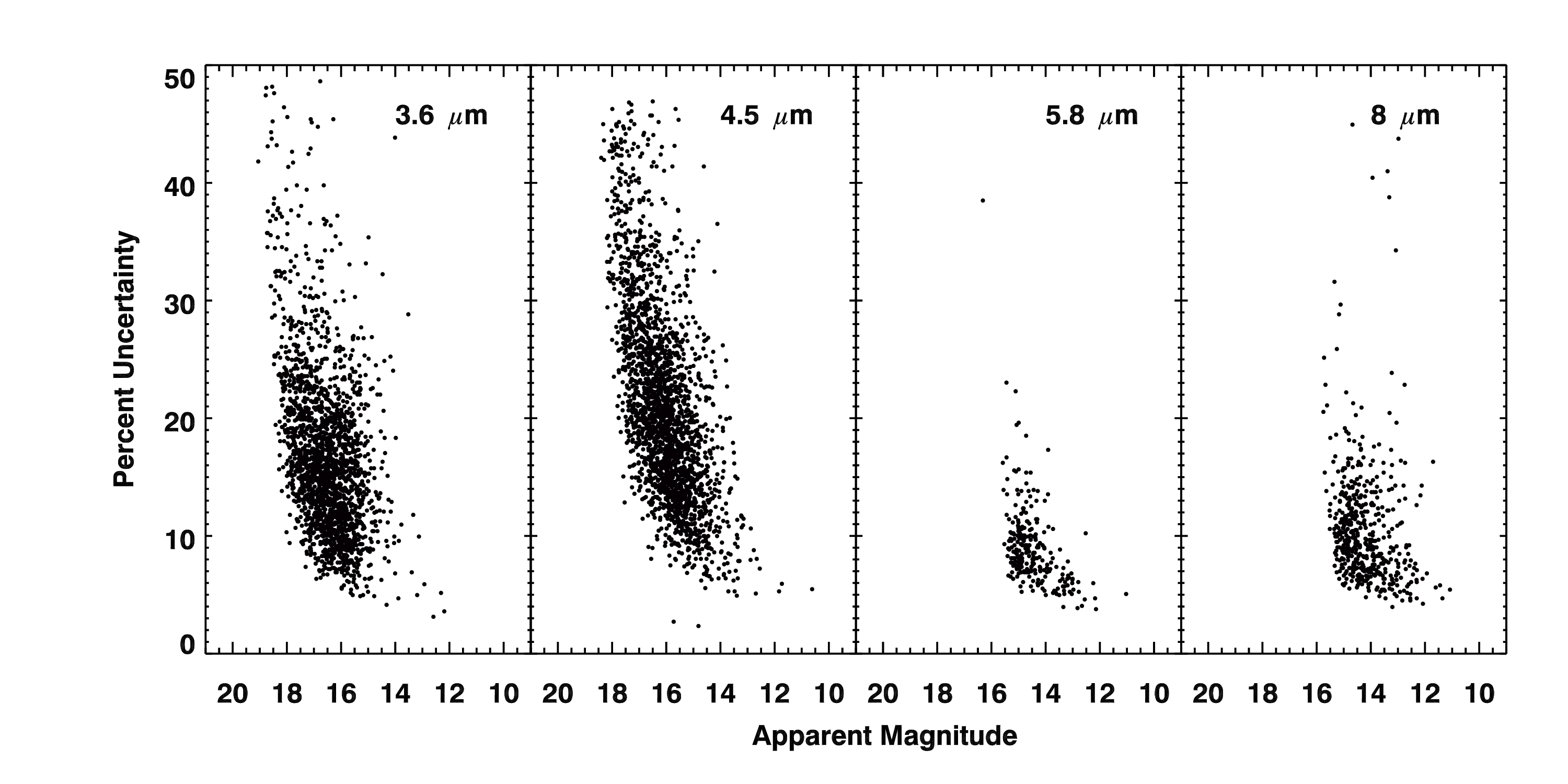}
\includegraphics[trim=0cm 0cm 0cm 2cm, clip=true,width=\textwidth]{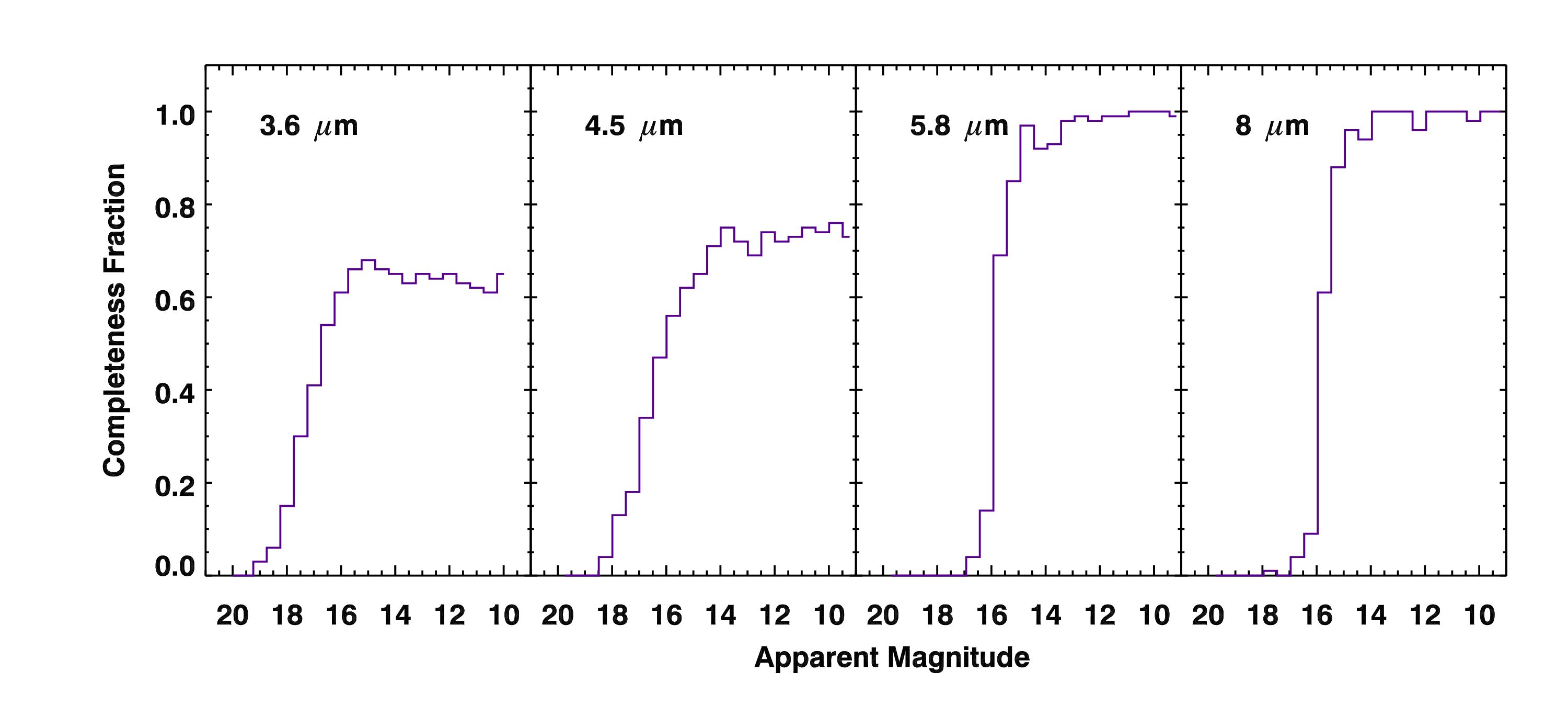}
 \caption[Photometric Completeness]{Photometric uncertainty and completeness fraction as a function of apparent magnitude for each IRAC band. The one-dimensional completeness fraction as a function of magnitude is shown in the bottom panels. Limits are averaged over the galaxy: photometry is more complete in the outskirts of the galaxy compared to the core.}
 \label{fig:completeLim}
\end{figure*}

The flux densities and uncertainties are colour--corrected using a 5000 K blackbody, according to the method described in the IRAC Data Handbook, version 3.0\footnote{\url{http://ssc.spitzer.caltech.edu/irac/dh/iracdatahandbook3.0.pdf}}. 
 Additionally, a pixel-phase-dependent correction \citep{Reach2005} was applied to the 3.6-$\mu$m photometry. Finally, magnitudes relative to Vega were derived using the zero-magnitude flux calibrations provided in the {\em Spitzer} IRAC Data Handbook. Figure~\ref{fig:completeLim} shows the representative photometric uncertainty as a function of source magnitude.  The IRAC instrument also has an absolute calibration uncertainty of 3 per cent \citep{Reach2005}, although we do not include this uncertainty in our photometric errors.

Table~\ref{tab:M32cat} lists the 1552 sources included in the final point-source catalogue. For a source to be included in the catalogue we require it to be detected in at least two IRAC channels; we consider a source to be a match if their centroids are within a 1\arcsec radius. The point sources detection statistics are summarised in Table~\ref{tab:M32catSummary}.

\begin{table}
\centering
\begin{minipage}{84mm}
 \caption{Number of point sources detected in M32}
 \label{tab:M32catSummary}
\centering
 \begin{tabular}{@{}cccccc@{}}
   \hline
   \hline
  Number           &     [3.6]  &    [4.5]   &	[5.8]   &  [8.0]  &  Total  \\
\hline  
 All detections    &    2177    &    2457    &    380   &    575  &  5589  \\  
 Final Catalogue   &    1270    &    1408    &    349   &    468  &  1552  \\
  \hline
 \end{tabular}
\end{minipage}
\end{table}


Completeness limits for each wavelength were determined via the extraction of artificial inserted stars. These false stars were placed at random pixel locations and have a limiting magnitude $\sim$ 2 mag fainter than the faintest objects recovered from our images. To avoid a significant increase in the source density of the field we limit the fraction of artificial stars introduced per image to $\sim$ 5 per cent of the total point sources detected. PSF-fitting photometry was then performed on the modified image. This process was repeated multiple times for each image to statistically analyse the photometric completeness of each band. Artificial stars are considered to be recovered if they are within a one-pixel radius of the input position and their magnitude differs by $|\delta m| \leq 1$ from their input magnitudes. If the magnitude difference is greater than this we consider the sources to be an unresolved blend of two or more stars and hence we do not consider them as recovered. The bottom panel in Figure~\ref{fig:completeLim} compares the fraction of injected and recovered stars for each IRAC band.

The different stellar density conditions found across our field of view have a large impact on the fraction of point sources recovered. Severe stellar crowding in the central regions of M32 results in a high degree of blending; this proves to be the major source of incompleteness. Furthermore, for  [3.6] and [4.5] the completeness factor of recovered stars in a given magnitude bin is much lower than at [5.8] and [8.0].  At shorter wavelengths, the increased sensitivity and brighter stellar emission results in a higher detection rate and thus a higher fraction of blended sources in crowded regions.


\begin{figure}
 \centering
\includegraphics[trim=2cm 0cm 0cm 0cm, clip=true,width=84mm]{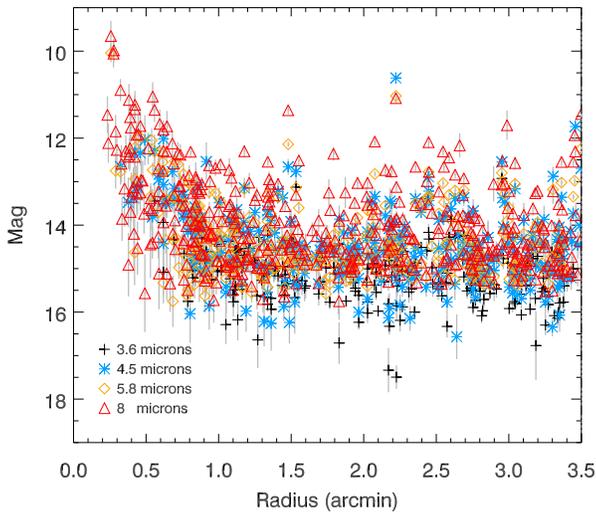}
 \caption[Apparent magnitude versus radial distance]{Distribution of the apparent magnitude with radial distance for all point sources detected at 8.0 $\mu$m. Towards the nucleus of M32 crowding is severe; sources with a magnitude enhancements in the core regions are probable blends.}
  \label{fig:Blendcheck}
\end{figure}

In regions of high surface brightness, unresolved blended sources in our catalogue will show a magnitude enhancement compared to individual point sources. 
Figure~\ref{fig:Blendcheck} plots the apparent magnitude as a function of radial distance. The relatively flat profile indicates that blended sources form only a minor component of our catalogue. Any sources in the inner regions that show a significant magnitude enhancement ($>5\sigma$ above the mean at $R >$ 1.5~\arcmin) are removed from our catalogue.

\section{Correcting star count numbers}\label{sec:starcount}

\subsection{M32 field contamination}

Typically, Galactic foreground stars as well as unresolved background galaxies will contribute to the source contamination in our catalogue. These contaminants will have a spatially uniform distribution across our field of view. To estimate the degree of contamination from foreground sources we use the {\sc trilegal} population synthesis code of \cite{Girardi2005} and we extrapolate the extragalactic background contamination from \cite{Fazio2004b}. In our IRAC field, we find that the contamination from foreground Galactic stars accounts for $\sim$ 6 per cent of the point sources detected at 3.6 $\mu$m in our photometric catalogue. Comparatively, the upper limit for the number of background galaxies is $\sim$ 18 per cent of objects detected at 8 $\mu$m. However, the final number of background galaxies included in our catalogue may be considerably lower as our point source extraction method rejects extended non-stellar objects.  An alternative estimate of the foreground star and background galaxy contamination may be obtained via colour-magnitude diagrams (Section~\ref{sec:M32CMDs}).

M32 has a projected separation of 24$^{\prime}$ (5.4 kpc) from M31's centre and is seen against the disc of M31.  Disc and halo stars from M31 are the dominant source of contamination in our M32 field. Furthermore, stars belonging to the nearby M31 galaxy will have similar IR colours to the evolved stellar population of M32. This makes it extremely difficult to isolate the individual stars belonging to M32, and the presence of some intruders is unavoidable. Combined with crowding, contamination from M31 is the most important limitation in the analysis of M32's rich stellar field. 

As contamination from M31 will not be homogeneously distributed across the field of view, the best estimate of the fraction of contaminating objects would be from a suitable control field located at the same isophotal level in the outer disc of M31 but away from M32's nucleus. Unfortunately, archival {\em Spitzer} IRAC observations of M31 and its surroundings only mapped the outer regions of M31 for about 120 seconds per sky position \citep{Barmby2006, Mould2008}. These observations do not provide sufficient depth to serve as a control field. In the following subsections, we describe the methods taken to mitigate the impact of M31 disc stars on our sample.

\subsection{Source density profiles}\label{sec:kingprofile}

\begin{figure}
 \centering
\includegraphics[trim=1cm 0cm 0cm 0cm, clip=true, width=84mm]{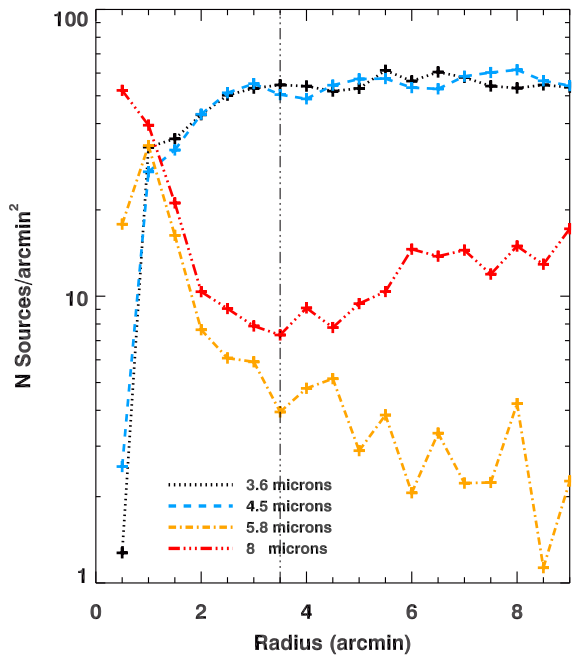}
 \caption[M32 Source Density]{Source density profiles for M32. {\em Spitzer} resolved the bright stellar populations in the outer halo of M32; however, the extremely high surface brightness and crowding in the bulge hinders detections in the central regions. The dashed line marks the radial limit where we have coverage in all four IRAC bands.} 
  \label{fig:SourceDensity}
\end{figure}

Using the radial density profiles we attempt to measure the IR contamination from M31 and Galactic foreground stars in each field of view.
To measure the stellar radial density profile of M32 we counted the sources contained in concentric annuli from its centre (where the bulk stellar population peaks) in steps of 0.5\arcmin. The number of sources per unit area was then obtained by dividing this value by the corresponding area covered by the annuli after correcting for regions not covered by our observations.
Figure~\ref{fig:SourceDensity} shows the source density profiles for M32.

In most channels the source density declines significantly toward the centre of M32, where incompleteness due to crowding effects in the core of M32 becomes significant. Both the 5.8- and 8-$\mu$m bands show a smooth decline in source density beyond a radius of $ R >$ 1.0\arcmin, until contamination from field stars and M31 begins to dominate the source density. This is particularly noticeable in the 8-$\mu$m band beyond $R >$ 4.5\arcmin, where there is a steady rise in source counts towards M31. 

The flattening of the star counts at $R =$ 2.5~--~4.0\arcmin~ was used to estimate the contamination from M31 and foreground stars. This corresponds to a point-source density of $\sim$5 sources per arcmin$^2$ for the 5.8 $\mu$m band and $\sim$7 sources per arcmin$^2$ for the 8 $\mu$m band. 
The flat stellar density profiles in the IRAC 3.6 and 4.5 $\mu$m bands beyond $R >$ 2.0\arcmin~ indicates a high degree of point source confusion with M31. Because of these high stellar surface densities we exclude these channels from the M32 stellar population analysis in Section~\ref{sec:lumFn}.


\begin{figure}
 \centering
\includegraphics[trim=1cm 0cm 0cm 0cm, clip=true, width=84mm]{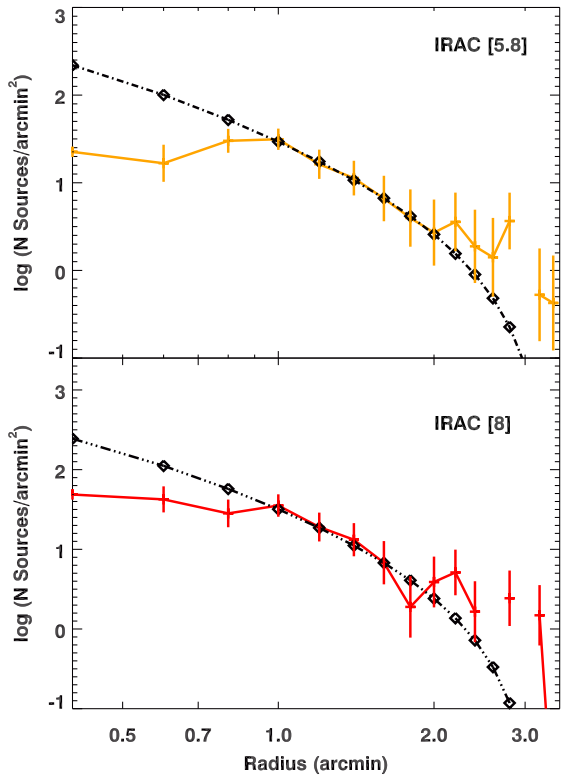}
 \caption[M32 King Profile]{King profiles fit to the background corrected IRAC 5.8 and 8 micron density profiles for M32. The data deviates from the King profiles due to source confusion inside 1 arc-minute. This indicates that we are underestimating the stellar density in the centre of M32.}
  \label{fig:KingProfileFit}
\end{figure}

To determine the radius at which our data becomes confusion limited we fitted the radial profiles of M32 with an empirical \citet{King1962} profile of the form:
\begin{equation}
I(R)=I_0\left[(1+\left({R\over R_c}\right)^2)^{-0.5}-(1+\left({R_t\over R_c}\right)^2)^{-0.5}\right]^2 \quad\,;\,R\le R_t
\end{equation}
where $I(R)$ is the density profile at projected radius $R$, $I_0$ is a normalising factor, $R_c$ the core radius, and $R_t$ is the photometric tidal cut-off radius. The best-fit King profiles to the background-corrected IRAC 5.8- and 8-$\mu$m density profiles are shown in Figure~\ref{fig:KingProfileFit}.  In both channels the data deviate from a King profile and become confusion limited at $R <$ 1\arcmin, which prevents us from studying the stellar populations in the innermost part of M32. At larger radial distances ($R >$ 2.5\arcmin), uncertainties in the background source density causes deviations from the profile shape. This also limits how well we can constrain the tidal radii. Table~\ref{tab:kingFit} lists the best-fit parameters. 

Due to the flat stellar-density profiles in the IRAC 3.6- and 4.5-$\mu$m bands we were unable to determine the corresponding King profiles for these channels. These flat profiles arise because of the low number of reliable detections within 90 arcseconds of M32's core and contamination from M31's disc at large radii where there is limited radial coverage for all angles.

\begin{table}
\centering
\begin{minipage}{85mm}
 \caption{King profile function best-fit parameters}
 \label{tab:kingFit}
\centering
 \begin{tabular}{ccc}
   \hline
   \hline
Parameter   &   IRAC 5.8  $\mu$m     &   IRAC 8.0  $\mu$m   \\          
\hline             
$I_0$               &  1103.68      &  1236.96    \\        
$R_c$  (arcmin)     &  0.241        &  0.245      \\                
$R_t$  (arcmin)     &  3.36         &  3.16       \\                
  \hline
 \end{tabular}
\end{minipage}
\end{table}


Based on the integrated surface brightness profile at optical wavelengths, \cite{Choi2002} find that the inner regions of M32 ($r < 150\arcsec$; 0.6 kpc) are predominantly spherical and well characterised by a de Vaucouleurs profile, while the outer regions show evidence of a residual elongated disc. We find a similar break at $r \approx 150\arcsec$, however, at large radii the number density of point sources per unit area becomes progressively more uncertain, and so our star-count analysis is of limited use for assessing a potential disc population. 

At M32's distance, the tidal radii in both the 5.8- and 8-$\mu$m bands correspond to a projected distance of 0.7 kpc. This value is somewhat lower than that given by \cite{Choi2002} of $R_t$ = 310$\arcsec$ (1.2 kpc), suggesting that the mid-IR stellar population maybe slightly more centrally concentrated than the underlying unresolved optical sources. This would be consistent with a more centrally concentrated population forming at the centre of M32's gravitational potential well, where gas would naturally collect. Such a distribution in the stellar population is consistent with the idea that the progenitor of M32 was a larger disc galaxy striped by interaction with M31. This scenario was proposed by \cite{Bekki2001} to explain M32's relatively high metallicity, its low gas content and the lack of a globular cluster system. However, it is also consistent with the bright mid-IR population arising in a more metal rich or younger population, both of which can be plausibly more centrally concentrated, without invoking mergers.

\subsection{Stellar spatial distribution}

Due to the limited spatial coverage, star counts measured in the off-field regions were also used to estimate the M32 field contamination. These regions (shown in Figure~\ref{fig:M32_obs_cov}) were obtained in conjunction with our primary observation of M32 and thus have the same depth as the M32 field. Star counts were measured using concentric annuli to sample a sub-section of each field, and cover the same solid angle as those used to measure the source density profile of M32. The off-field annuli centred near $\alpha = $0:42:56.10, $\delta = +$40:45:48.38 sampled the outer disc of M31 where we have coverage at 3.6 and 5.8 $\mu$m. The other set of off-field annuli was centred at  $\alpha = $0:42:27.75, $\delta = +$40:57:48.76 in the field closer to the main body of M31 observed at 4.5 and 8 $\mu$m. In these regions the stellar density of M31 should average out across the field of view, producing a flat density profile.

Located furthest away from M31, the IRAC 5.8-$\mu$m field suffered the least contamination, with an average background point-source density of 2.5 sources per arcmin$^2$, while the 8-$\mu$m field (which intersects the inner disc of M31) has an average density of 11 sources per arcmin$^2$. 
Based on the ratios of sources in the off-fields, we expect  $\lesssim$ 20 per cent of the point sources in the on-field region to be contaminants from M31 in the 5.8- and 8-\mum bands. This assumes that objects associated with M32 do not contribute to the number density in the off-field regions.

\section{Results \& Discussion}\label{sec:M32results}

\subsection{M32 luminosity functions}\label{sec:lumFn}

In this section, we examine the mid-IR stellar luminosity functions of M32. These are useful in constraining the star-formation history and provide an estimate of the dust emission.
The luminosity functions for each of the IRAC bands were constructed by sorting the point sources into bins based on their magnitude. The bin size of the luminosity functions is set to 0.2 mag, to account for photometric errors in our catalogue.  To reduce the effects of contamination by M31 as much as possible, while still retaining a reasonable sample size, only sources which are detected in at least two wavebands and that are within 3.5\arcmin~ of M32's centre were counted. It should also be noted that the luminosity function at 3.6 $\mu$m will be suppressed due to the extreme crowding in the centre of the galaxy. Furthermore, if the centre contains a different stellar population than the outer part of the galaxy, the shape may also be affected.

Comparisons of the 3.6-\mum luminosity function with and without stars in the central 1\arcmin~ yield similar results; this is not unexpected as the increase in crowding near the core limits the number of stars that can be reliably resolved. At 8 $\mu$m the situation is different. At these wavelengths crowding has less of an impact and we are sensitive to sources with $r > 0.6\arcmin$ of the core. Removing sources within  $r < 1\arcmin$ effectively scales the luminosity function downwards and we see a very slight change in the gradient. At the faint end of the luminosity function we detect very few sources within 1\arcmin; it is here where the effects of crowding and the high surface brightness has some impact.

The uncertainties in the luminosity functions are set both by Poisson statistics of the sample and by the uncertainties in the background counts for each magnitude bin.  Defining a complete, uncontaminated sample consisting solely of M32 sources is not possible with currently available data; thus for the present survey, the luminosity functions presented here provide an upper limit to the true luminosity distribution, as some contamination may be present in each magnitude bin.


\begin{figure*} 
\centering
\includegraphics[trim=1cm 0cm 0cm 0cm, clip=true,width=\textwidth]{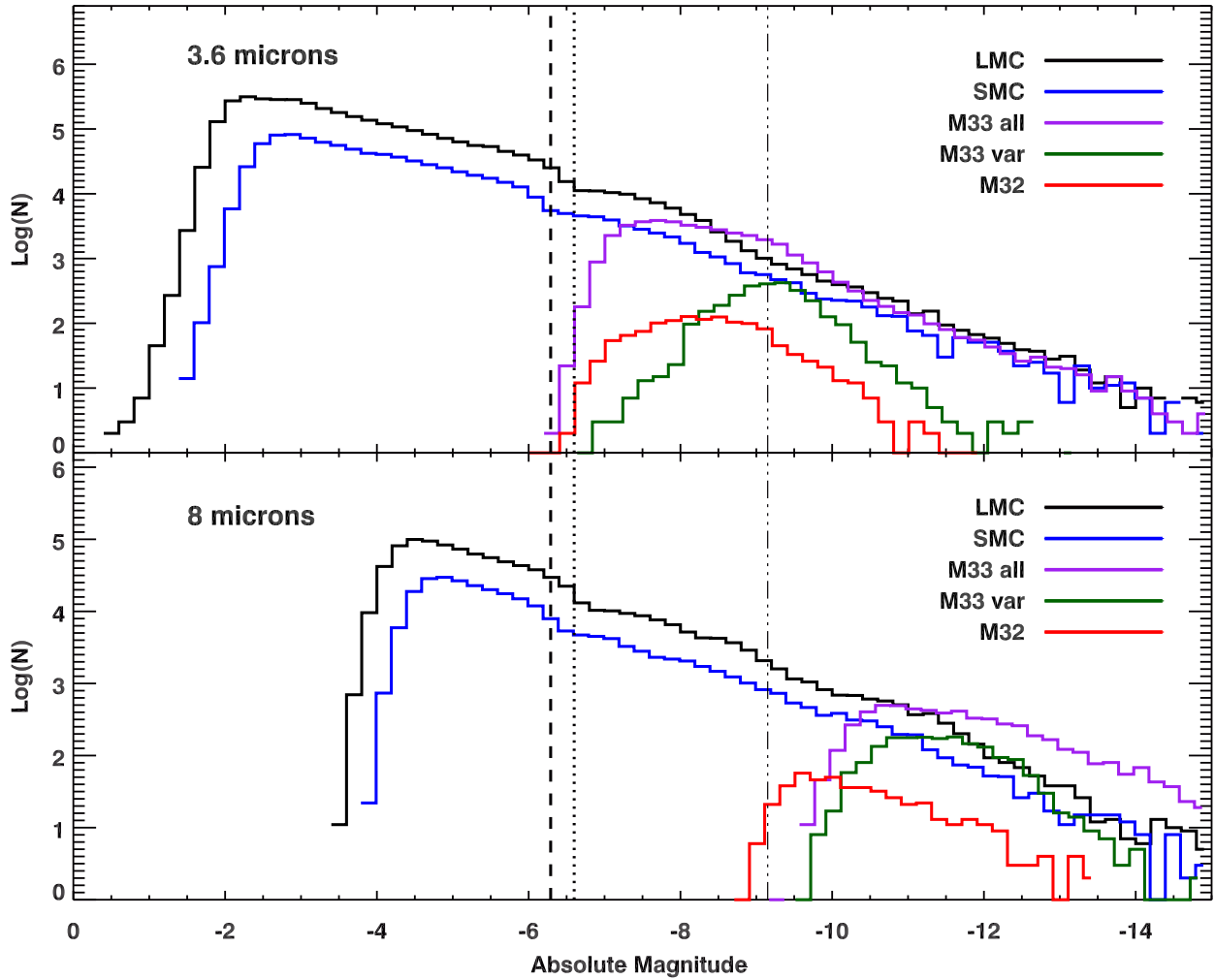}
 \caption[Comparison of the M32 and Magellanic Cloud luminosity functions]{Comparison of the M32, M33 and Magellanic Cloud luminosity functions.  The dotted-dashed line marks the 50\% completeness limit. The dotted line marks the 3.6 \mum TRGB for the LMC and the dashed line shows the TRGB for the SMC. } 
  \label{fig:M32_MC_LuminosityFunction}
\end{figure*}

\begin{figure*} 
\centering
\includegraphics[trim=1cm 0cm 0cm 0cm, clip=true,width=\textwidth]{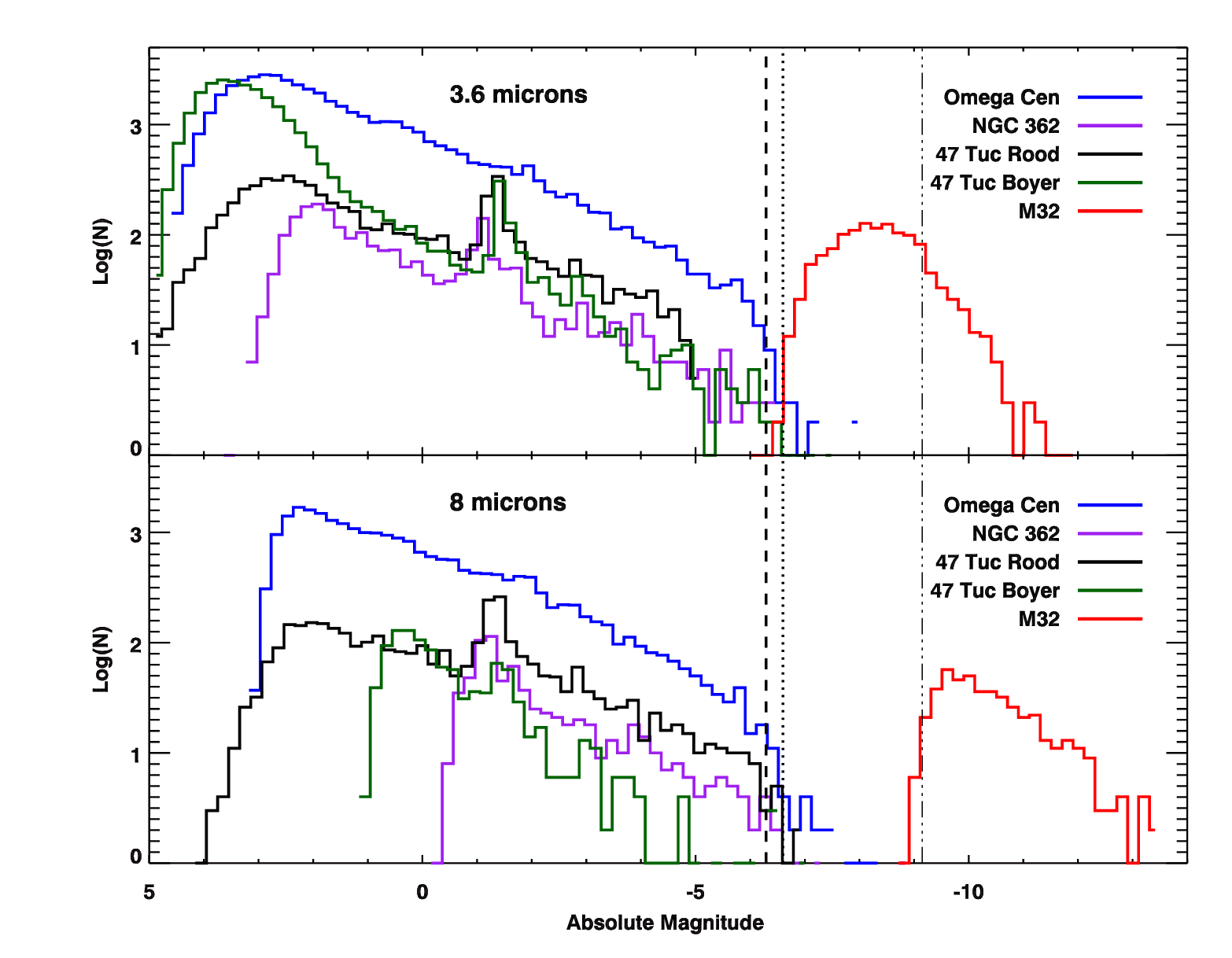}
 \caption[Comparison of the M32 and 47 Tuc luminosity functions]{Comparison of the luminosity functions of M32 and 47 Tuc. The red line represents all sources detected within 3.5\arcmin of M32's centre; the black line represents the IRAC deep `Rood' observations of 47 Tuc \citep{Origlia2007, Boyer2010} and the green line represents the SAGE-SMC observations of 47 Tuc \citep{Boyer2010}. This indicates that the recovered stars in M32 are significantly brighter than the TRGB.}
  \label{fig:47Tuc_Luminosity Function}
\end{figure*}

Figure~\ref{fig:M32_MC_LuminosityFunction} shows the {\em Spitzer} IRAC luminosity functions at 3.6 and 8.0 \mum for M32. The profile shape at 3.6 \mum is similar to the variable stars in M33 \citep{McQuinn2007}, but it has a steeper falloff at bright magnitudes compared to all detections in the Magellanic Clouds. This is likely due to a difference in the metallicity of the stellar populations and hence the ratio between C-rich and M-type AGB stars \citep{Mouhcine2003, Battinelli2005, Cioni2009}.

At 8.0 \mum the luminosity function drops smoothly with magnitude and is similar in appearance to both the Magellanic Clouds and M33; at this wavelength the luminosity function probes the dust emission rather than the photospheric temperature. The large number of point sources in M32 indicates a significant population of dust-producing stars.

The tip of the AGB can be estimated from the luminosity functions by identifying the magnitude where the source count decreases significantly. Although source counts are low, an inspection of the luminosity functions suggests that the tip of the AGB for M32 is located near  M$_{3.6}$ $\sim -10.8 \pm$ 0.6 mag. M32 candidate stars brighter than the tip of the AGB are either luminous RSGs in M31 or foreground sources. The position of the tip of the AGB indicates that the IR-stellar population of M32 appears older than the Magellanic Clouds, which have experienced an extended star forming history. 
The location of the tip of the red giant branch (TRGB) is relatively insensitive to the metallicity of the host population \citep{Cioni2003}, and lies at  M$_{3.6}$ $\sim -6$ mag, below the completeness limit of our survey. We therefore assume that the majority of the point sources detected in our catalogue are intermediate-age AGB stars, with some small contamination by other stellar types.

In contrast to the luminosity function of globular clusters where there is only a sparse population of AGB stars above the TRGB (Figure~\ref{fig:47Tuc_Luminosity Function}), M32's luminosity functions are strongly dominated by a bright evolved stellar population. 
Stars in Galactic globular clusters are typically found to be 10-14 Gyr old \citep[e.g.][]{Roediger2014}, while bright carbon stars are predicted to be much younger; with ages in the region of 8 Gyr to 100 Myr, depending on the mass and metallicity of the system \citep{Karakas2007, Marigo2007}.
The majority of stars detected at 8 $\mu$m in M32 are thus much younger than the Population II stars in globular clusters, and the numbers suggest we are tracing a metal-rich population of intermediate age. 

We detect a large number of dusty stars compared to the Local Group dwarf irregular galaxies WLM, IC 1613, Phoenix, LGS 3, DDO 210, Leo A, Pegasus dIrr, and Sextans A \citep{Jackson2007a, Jackson2007b, Boyer2009} commensurate with the greater luminosity of M32. These composite stellar populations span a wide range in age and metallicity, however, any further comparison is limited due to the low numbers of stars in these systems.

\subsection{IRAC Colour distribution}\label{sec:M32colDist}

\begin{figure} 
\centering
\includegraphics[width=84mm]{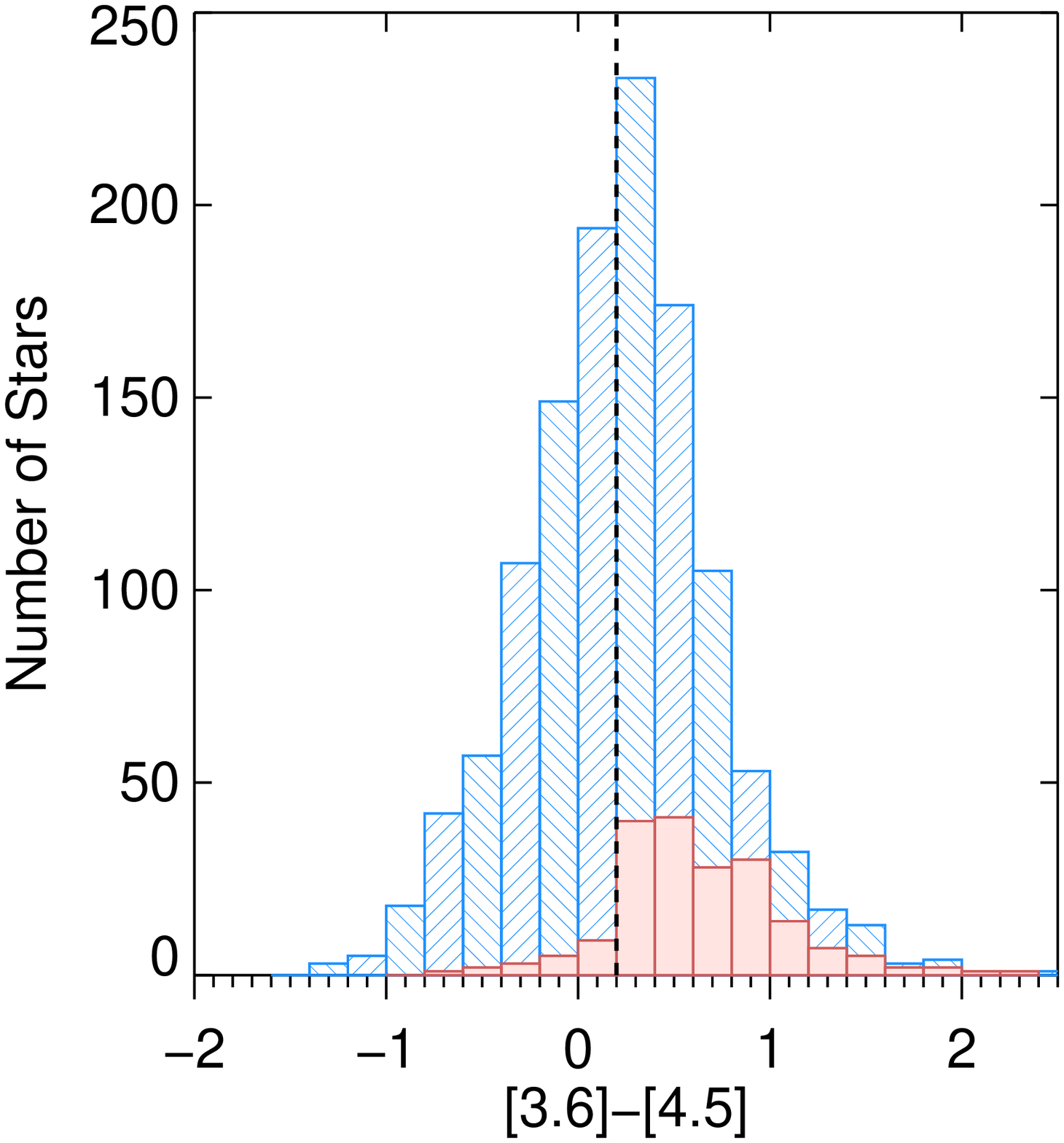}
 \caption{The IRAC $[3.6]-[4.5]$ colour distribution of all detected sources in M32, shown in blue. Objects which have a measured [8.0] mag in M32 are denoted by red. Stars detected at [8.0] and appearing to the right of the dashed line are probably carbon-rich.}
  \label{fig:M32_HistColourDist_34_N}
\end{figure}

\begin{figure} 
\centering
\includegraphics[width=84mm]{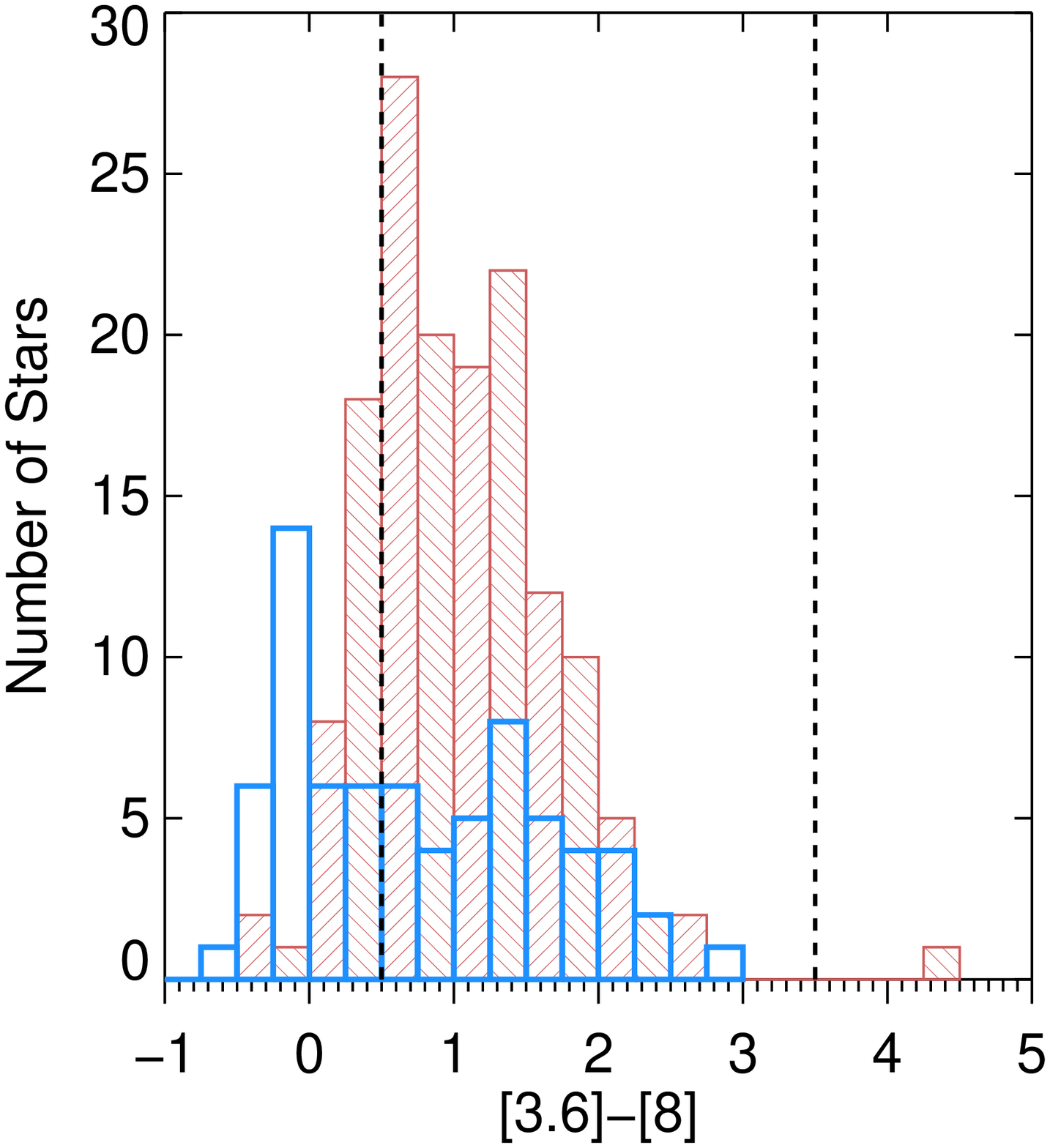}
 \caption{The IRAC $[3.6]-[8.0]$ colour distribution of sources in M32. Bright sources with  M$_{3.6} < -9.5$ are shown in blue while faint sources (M$_{3.6} > -9.5$) are shown in red. The dashes vertical lines mark the range used to select Carbon star candidates.}
  \label{fig:M32_HistColourDist_38_N_mag}
\end{figure}

Figures~\ref{fig:M32_HistColourDist_34_N} and \ref{fig:M32_HistColourDist_38_N_mag} show histograms of colours for point sources in our catalogue. Oxygen-rich giants typically have a slightly blue [3.6]--[4.5] colour distribution, due to photospheric SiO and CO absorption at 4.08 $\mu$m and 4.66 $\mu$m, respectively, while HCN and C$_2$H$_2$ absorption features at 3.5 $\mu$m and 3.8 $\mu$m, within the 3.6 \mum band cause carbon stars to appear red \citep{Marengo2007}. Foreground main-sequence dwarfs are also slightly reddened. In Figure~\ref{fig:M32_HistColourDist_34_N} the blue-shaded area shows all the sources detected at 3.6 and 4.5 $\mu$m; this population is a mix of foreground dwarf stars, contaminants from M31 and AGB stars in M32. 
Evolved stars undergoing significant mass loss become dust-enshrouded and exhibit a strong mid-IR excess. Consequently, objects detected at [8.0] can be tentatively separated into oxygen-rich and carbon stars; this restriction helps to minimise contamination from foreground sources. As the IRAC sensitivity limit for the 8-$\mu$m data is brighter than at 3.6 and 4.5 $\mu$m: bright evolved AGB stars with strong circumstellar dust emission are preferentially detected over fainter, less-evolved AGB stars. This favours carbon stars and might explain the low number of oxygen-rich objects in our sample. 

Figure~\ref{fig:M32_HistColourDist_38_N_mag} separates the [3.6]--[8.0] colour distribution as a function of magnitude, following \cite{McQuinn2007}. The bright sources with M$_{3.6} < -9.5$ have a binomial distribution with peaks at $[3.6]-[8.0] \approx 0$  (which corresponds to AGB stars without dust) and $[3.6]-[8.0] \approx 1.5$ (sources reddened by circumstellar dust emission).  Based on the colour distribution of variable stars in M33, carbon star candidates are thought to fall in the range $0.5 \lesssim [3.6]-[8.0] \lesssim 2.5$. Other non-variable sources reddened by circumstellar dust emission, for example, young stellar objects would have $[3.6]-[8.0] \gtrsim 2.8$.

\subsection{IRAC Colour-Magnitude Diagrams}\label{sec:M32CMDs}
 
\begin{figure} 
\centering
\includegraphics[width=84mm]{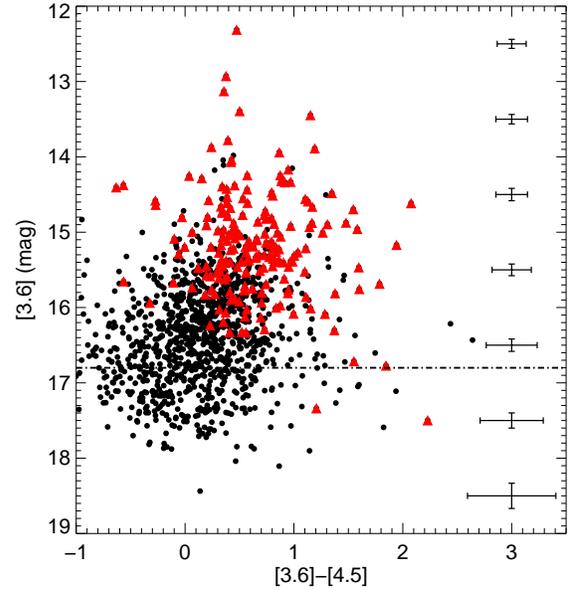} \\
\includegraphics[width=84mm]{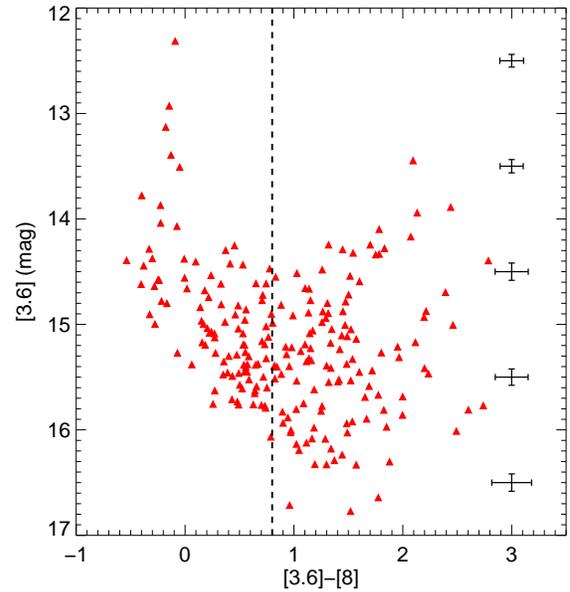} 
 \caption[IRAC 3.6 \mum CMDs of M32]{{\em Spitzer} IRAC CMDs of M32. For all panels the y-axis is the apparent 3.6 \mum magnitude. The red triangles indicate a 8-\mum counterpart. The initial colour-cut used to select x-AGB candidates is indicated by the vertical dashed line, while the horizontal line represents the theoretical TRGB for M32. The error bars show representative uncertainties; they do not vary significantly with colour.}
  \label{fig:M32_CMD_3}
\end{figure}

\begin{figure} 
\centering
\includegraphics[width=84mm]{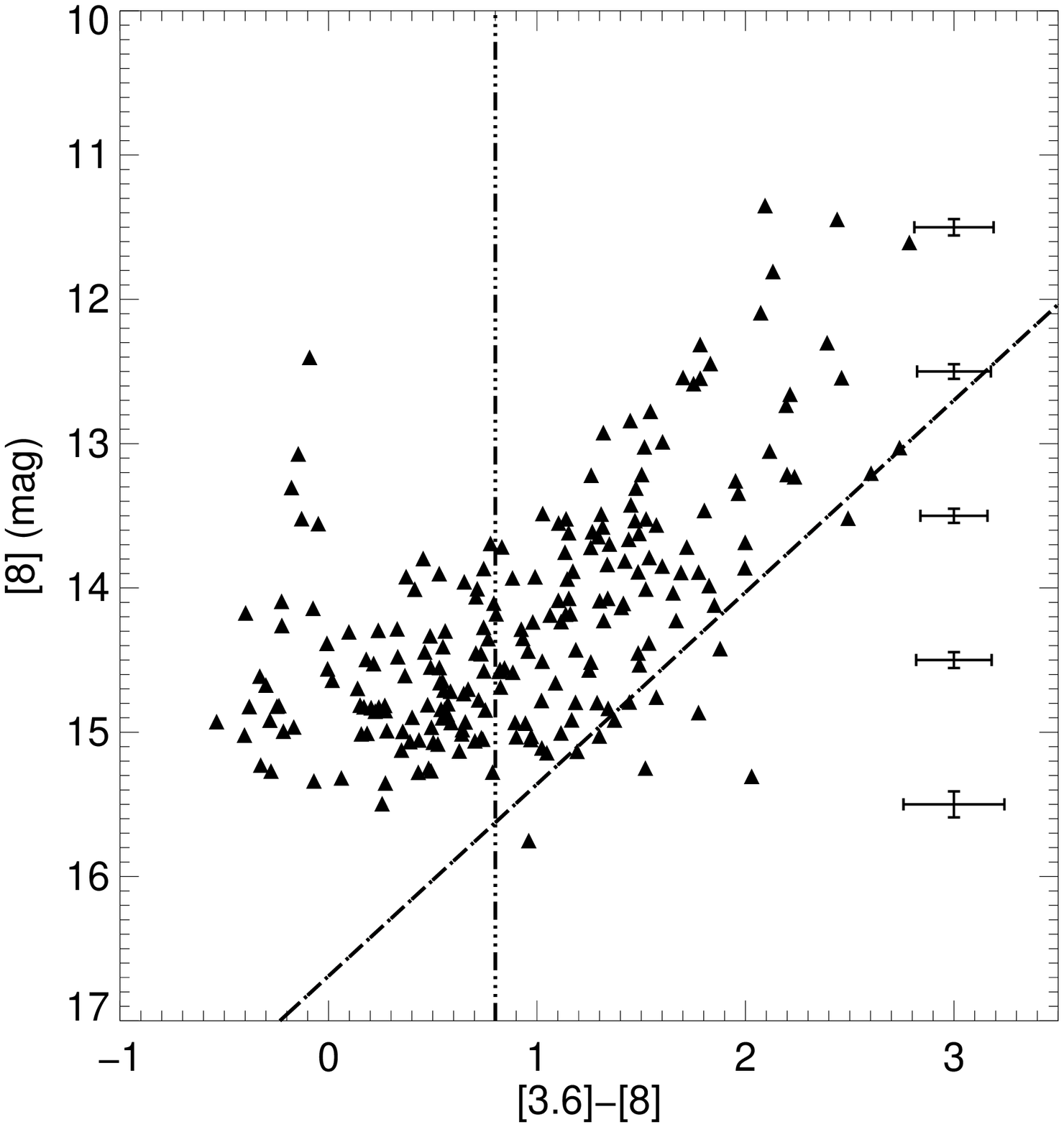} \\
\includegraphics[width=84mm]{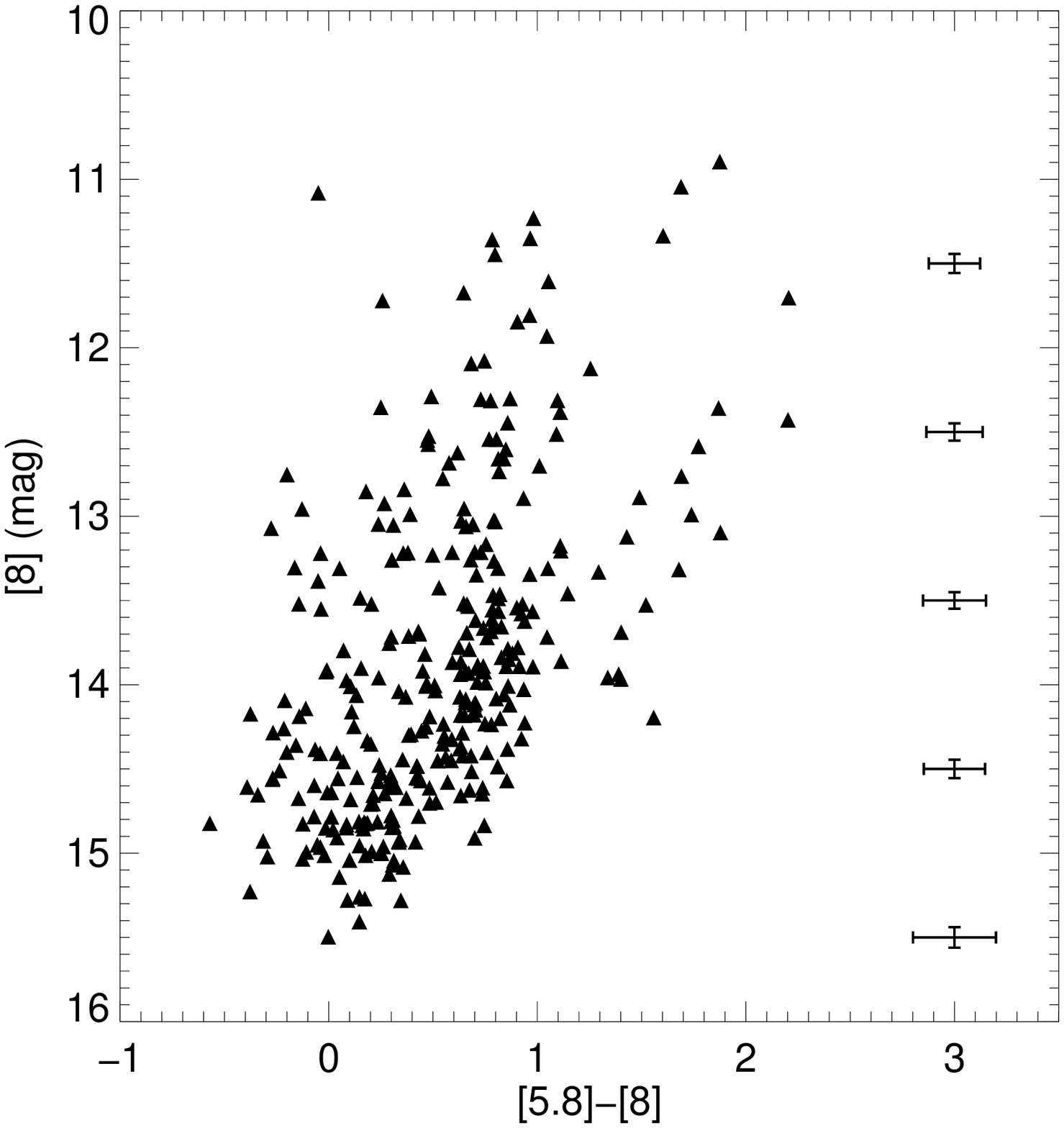} 
\caption[IRAC 8.0 \mum CMDs of M32]{{\em Spitzer} IRAC CMDs of M32. For all panels the y-axis is the apparent 8.0 \mum magnitude. The colour-cuts used to select x-AGB candidates are indicated by the dash-dot lines lines. The error bars show representative uncertainties; they do not vary significantly with colour.}
  \label{fig:M32_CMD_8}
\end{figure}

Mid-IR colour-magnitude diagrams (CMDs) for all catalogue point sources are presented in Figures~\ref{fig:M32_CMD_3} and \ref{fig:M32_CMD_8}. At these wavelengths stellar temperature no longer affects the CMDs (as the SED is described by the Rayleigh-Jeans tail), and any features seen are due to the absorption and emission by circumstellar molecular and dust species.
The well-defined vertical branch with zero colour ([3.6]--[8.0] $\approx$ 0) in these diagrams traces foreground stars and dust-free blue-supergiants in M31. The relatively low number of sources that fall on this branch indicates a nominal level of contamination from foreground sources within our field of view.

Based on the maximum luminosity at 3.6 \mum and the slightly red vertical [3.6]--[4.5] colour distribution, the foreground sources are probably main-sequence dwarfs. Unlike giants, dwarf stars have water vapour absorption features at 3.6 \mum and lack CO absorption at 4.5 \mum resulting in a shift towards the red \citep{Merrill1976}; foreground giant stars would also appear brighter.

As we probe to fainter magnitudes, photometric errors become more significant. This may be responsible for the large spread in colour for  [3.6] $\ge$ 16.5 in the [3.6] vs.~[3.6]--[4.5] colour-magnitude diagram (Figure~\ref{fig:M32_CMD_3}, top panel). 
It should be noted that sources towards the centre of M32 may be not accurately extracted at one or more wavelengths due to crowding. Although care had been taken to remove blended sources from the catalogue, blends may potentially affect the derived colours by enhancing the IRAC flux. We expect this to influence $\lesssim$ 4 per cent of the sources in the CMDs, if we assume all detections with magnitude enhancements greater than 3$\sigma$ are blends. 

The [8.0] vs.~[5.8]--[8.0] colour-magnitude diagram (Figure~\ref{fig:M32_CMD_8}, bottom panel) provides the best representation of the `true' stellar population of M32, as these bands are least affected by contamination from M31.

The stellar sequences traced by the diagonal branches in the CMDs are difficult to isolate, making it harder to accurately classify the different populations of cool evolved stars. Using the mid-IR colour-classification scheme from \cite{Boyer2011}, we identify a large population of  dust-enshrouded evolved stars (x-AGB stars) with  [3.6]$-$[8] $>$ 0.8. This corresponds to 55 per cent of the sources with detections in both channels. 
In non-star-forming galaxies like M32 contamination from young stellar objects should be negligible; however unresolved background galaxies may contaminate the x-AGB sample. Contamination from unresolved background galaxies may be reduced by requiring x-AGB sources to be brighter than an empirical boundary set for extra-galactic objects. Thus we apply an additional colour-cut in the [3.6]$-$[8] vs.~[8] CMD, which is adjusted for M32's distance from the classification by \cite{Boyer2011} and defined as: 
\begin{equation} [8] = 17.4 - 1.33 \times ([3.6] - [8]) ; \, \, \,\, \,   [3.6] - [8] < 3.0. \end{equation}
There are 11 sources that do not fall within this range, reducing the final number of x-AGB candidates to 110. The colour-cuts used to identify the x-AGB candidates are plotted in Figures~\ref{fig:M32_CMD_3} and \ref{fig:M32_CMD_8}. Although an effort was made to remove contaminating sources from the final list of the x-AGB candidates, we still expect a moderate contamination from background galaxies. Other colour-cuts used to identify O-AGB and C-AGB stars require additional near-IR photometry.


The x-AGB sample is thought to be dominated by carbon stars \citep{vanLoon2006, Boyer2012}, although some cross-contamination by oxygen-rich AGB stars is expected. At LMC and SMC metallicities, 97 per cent of the x-AGB candidates were matched to a carbon-rich {\sc grams} model \citep{Riebel2012}. Spectroscopic observations of the Magellanic Clouds confirm this: less that 10 per cent of the candidate x-AGB stars were identified as oxygen-rich stars \citep{Woods2012, JonesPhD}. 
Previous estimates of M32's metallicity suggest that the galaxy is predominantly metal rich, with a near-solar [Fe/H] value which trails off to [Fe/H]$\sim$--1. Consequently, we expect to see a larger ($>4$ per cent) fractional contribution from oxygen-rich AGB stars in the x-AGB class \citep{Boyer2013}, although without spectroscopic confirmation how much more is unclear.  Due to sensitivity limits these O-rich stars are likely to be more massive than 4 \Msun~and have undergone hot-bottom burning for the star to remain O-rich. 

At metallicities comparable to M32, AGB stars will become carbon-rich if their initial mass is in the region of $~$1.2~--~3.4~M$_{\odot}$ \citep[figure 9]{Straniero1997, Marigo2007,McDonald2012}; this corresponds to a stellar age between 0.2 to 5 Gyr. This boundary is dependent on the choice of evolutionary model and may potentially extend to slightly higher ages (up to 8 Gyr). Thus the brightest x-AGB stars may be the progeny of a younger subset of the stellar population of M32. This agrees well with suggestions that the metal-rich component of M32 was formed during a period of star formation that occurred less than 7 Gyr ago \citep{Monachesi2011}. 

Since our data is more sensitive to carbon stars which trace a younger stellar population than the majority of oxygen-rich AGB stars (which have a lower initial mass), we cannot exclude the possibility that an older oxygen-rich population is also present in M32. 
 Evidence for multiple star-formation events in M32 has been presented by \cite{Monachesi2012} using stars near the main sequence turnoff. These multiple star-formation events may have been induced by tidal interactions with M31 or through accretion of gaseous material \citep{Bekki2001, Block2006, Hammer2010}. An extended period of star-forming activity associated with a much larger disc-dominated galaxy would increase the numbers of old AGB stars; unfortunately these faint AGB stars fall below our observational limit.


\begin{figure} 
\centering
\includegraphics[width=84mm]{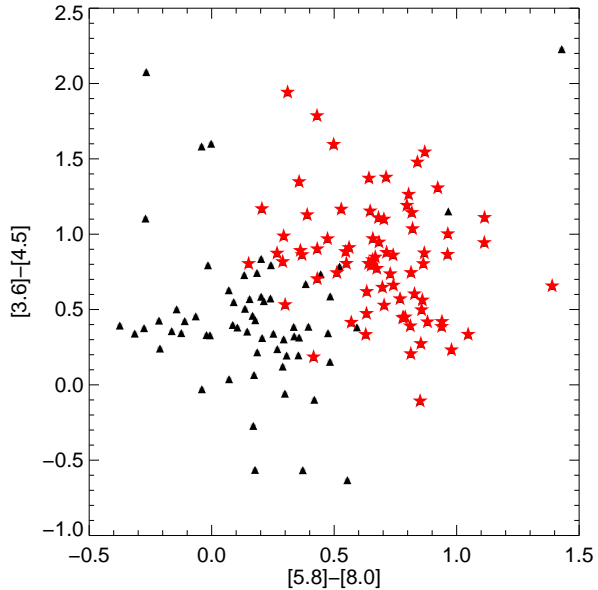}
 \caption[IRAC \lbrack3.6\rbrack--\lbrack4.5\rbrack~ vs.~\lbrack5.8\rbrack--\lbrack8.0\rbrack ]{IRAC $[3.6]-[4.5]$ vs.~$[5.8]-[8.0]$ colour-colour diagram, showing the 139 point sources identified at all bands in our M32 catalogue. The extreme AGB stars in our sample are indicated by red stars.}
  \label{fig:M32_CDD}
\end{figure}

In Figure~\ref{fig:M32_CDD} we show the IRAC $[3.6]-[4.5]$ vs.~$[5.8]-[8.0]$ colour-colour diagram of all the sources in our catalogue. The extreme AGB stars in our sample are indicated by red stars. These sources are often obscured in the optical and near-IR due to circumstellar dust extinction ($A_{\rm v} \approx 3-4$) and are bright at 8 $\mu$m due to thermal emission by the same dust. The red colours for the majority of the other sources are consistent with mass-losing AGB stars. 

\subsection{Mass-loss rates of the AGB candidates}\label{sec:M32MLRs}

Using mid-IR colours as a proxy for dust-production rates, we can estimate the total dust input from the AGB candidates in M32. We derived mass-loss rates using the dependence on [3.6]-[8.0] colour derived empirically by \cite{Matsuura2009}, for the 160 AGB candidates in our catalogue which have an IR-excess of $[3.6]-[8.0] > 0.5$ mag. This method of using only mid-IR colours to determine mass-loss rates for AGB stars relies on a number of simplifying assumptions; as such these values should be treated as a first-order estimate into the dust production in M32. We note that the prescription used to estimate the mass-loss rates assumes that the stars in our catalogue have a single  dust composition, luminosity, wind speed, effective temperature (T$_{\rm eff}$) and solar metallicity.  As the colour excess only provides a measurement of the dust content, a Galactic gas-to-dust ratio of 200 was adopted to convert the measured dust mass to a gas mass-loss rate.  This conversion factor is highly uncertain and it is unclear if this ratio is applicable to extra-Galactic evolved stars. Due to the large number of assumptions we have made about a source, the individual mass-loss rates for the candidate AGB stars have high uncertainties.  However, for statistically large samples cumulative mass-loss rates should provide a reasonable approximation. 

Individual stellar mass-loss rates range from $6 \times 10^{-8}$ to $2.6 \times 10^{-5}$ ${\rm M}_{\odot} \, {\rm yr}^{-1}$ and the cumulative mass-loss rate is $(1.5-1.8) \times 10^{-4}$ ${\rm M}_{\odot} \, {\rm yr}^{-1}$, depending on the fraction of sources that belong to the evolved stellar population. Background galaxies and other interlopers that occupy the same region of the CMDs as dusty evolved stars may result in some overestimates in the total dust input from sources in our catalogue. 
Dust production is dominated by the five most extreme sources in our catalogue, which produce over 30 per cent of the total dust input. The reddest source in our sample contributes 15 per cent of the total. However, this source falls outside the colour-cut for x-AGB stars and has a strong 24-\mum excess; it is probably an unresolved background galaxy, thus we exclude it from our total.

As the candidate x-AGB stars are centrally concentrated about M32 rather than distributed evenly across the IRAC field of view they are less likely to be background galaxies. 
The colour-derived, integrated mass-loss rate for the x-AGB candidates is $1.45 \times 10^{-4}$ ${\rm M}_{\odot} \, {\rm yr}^{-1}$, which corresponds to  97 per cent of the total dust input in the galaxy. The remaining 3 per cent is returned from AGB candidates with moderate IR excess. For comparison, we also estimate the cumulative mass-loss rate using the colour-MLR relation derived by \cite{Gullieuszik2012}; this was found to be $2.1 \times 10^{-4}$ ${\rm M}_{\odot} \, {\rm yr}^{-1}$.

Due to the severe stellar crowding in the core region of M32, not all the red AGB stars with extremely high mass-loss rates will be accounted for due to the confusion. Thus the cumulative mass-loss for M32 is likely to be a lower limit. 
Other evolved sources in our catalogue produce only moderate amounts of dust and gas (less than $10^{-6}$${\rm M}_{\odot} \, {\rm yr}^{-1}$) and contribute little to the integrated value. 


Overall, our findings for the total dust input agree with other recent studies. In the LMC ($Z = 0.5 Z_\odot$) the global dust injection rate from evolved stars was found to be on the order of $2 \times 10^{-5}$ ${\rm M}_{\odot} \, {\rm yr}^{-1}$, and for the SMC ($Z = 0.25 Z_\odot$) evolved stars produce $9 \times 10^{-7}$ ${\rm M}_{\odot} \, {\rm yr}^{-1}$ of dust \citep{Riebel2012, Boyer2012}. In both galaxies, extreme AGB stars are the primary contributors to the total dust production, accounting for approximately 65 per cent of the dust production but only 3 per cent of the sample by number.    
 Similarly, the dust input for several more metal-poor dwarf irregular galaxies in the Local Group was found to be governed by a few AGB stars undergoing intense mass-loss via superwind \citep{Boyer2009}. Here integrated total mass-loss rates range from $4.4 \times 10^{-5}$ to $1.4 \times 10^{-3}$ ${\rm M}_{\odot} \, {\rm yr}^{-1}$.


One notable difference between M32, the Magellanic Clouds and the dwarf irregular galaxies is the metallicity of the host population. In metal-poor environments, the dust yield from oxygen-rich AGB stars is reduced, unlike carbon stars where dust yields do not depend significantly on the initial metallicity of the progenitor \citep{Groenewegen2007, Sloan2008, Sloan2012}. Consequently the fractional contribution from oxygen-rich dust may be higher in metal-rich environments like M32.

\cite{Javadi2013} estimate dust-production rates for $\sim$110 variable stars detected at 8 \mum in the central square kiloparsec of M33 that are of near-solar metallicity. Despite there being a comparable number of stars detected at [8.0] in M33 as M32, the total mass-return in M33 is much higher due to the contribution of several massive, very dusty stars with mass-loss rates of a few $\times 10^{-5}$ ${\rm M}_{\odot} \, {\rm yr}^{-1}$. This estimate also includes red giant variables for which no mid-IR counterpart was identified.
In contrast to the Magellanic Clouds, the dust returned in the central regions of M33 from low-mass oxygen-rich AGB stars appears to be comparable to the contribution from the more massive carbon stars and red supergiants. However, due to the distance of M33 (which is not much different to that of M32) and the relatively sparse points that comprise the spectral energy distribution, there is a great deal of uncertainty in the classification into carbon and oxygen rich stars. Furthermore, like M32, the AGB population detected by {\em Spitzer} in M33 is dominated by candidate x-AGB stars with only a small fraction of non-extremes; thus there is sufficient information at this time to constrain which species of dust is present.

If a large fraction of the x-AGB stars in M32 turn out to be oxygen-rich then the dust return could be several times higher; this is due to the lower specific opacity of silicates compared to amorphous carbon grains. Owing to the higher metallicity of M32, O-rich stars might be expected to dominate the dust input at lower luminosities; unfortunately, the high surface brightness, limited sensitivity and resolution conspires against dust-injection rates being calculated for the less luminous (oxygen-rich) sources. Current estimates for the dust yield from non-extreme sources which comprise 25 per cent of stars with an IR excess, show that they return approximately 3 per cent of the measurable dust production. Hence, AGB stars significantly fainter than our detection limits are not expected to show any appreciable contribution to the dust-production rate in M32.

\subsection{Comparison to Davidge (2014)}

During the refereeing process, the publication of an independent reduction of this data-set was made public \citep{Davidge2014}. Comparisons of the mid-IR CMDs show sources extracted by \cite{Davidge2014} reach fainter magnitudes at than sources in our catalogue; this is due to differences in the level above background chosen for extraction and the stringent criteria we require for inclusion in our catalogue. This invariably means that some real point-sources are rejected from our sample, but contamination from blends, non-stellar objects, and background galaxies are reduced, enabling a more accurate estimate of the dust-production rate for M32.
\cite{Davidge2014} supplement the mid-IR IRAC data with near-IR ground based observations to compare the present-day stellar population of M32 with the star-forming history measured by \cite{Monachesi2011}. Although our methods of analysis differ we obtain similar estimates in age (on the order of a few gigayears) for the dusty evolved stars in M32.

\section{Summary and conclusions}\label{sec:conclusion}

We have presented mid-infrared observations of the dwarf elliptical galaxy M32, obtained with the Infrared Array Camera on board the {\em Spitzer Space Telescope}. These images resolve individual stars in the bulge of M32, revealing a rich population of dusty evolved stars. In Section~\ref{sec:kingprofile} it is shown that this population is more centrally concentrated than the underlying optical sources, indicating a radial population gradient.

Despite the strong levels of contamination from M31, we find that luminous stars with significant dust emission dominate the 5.8- and 8-$\mu$m luminosity functions. We estimate the tip of the AGB to be M$_{3.6}$ $\sim -10.8 $ mag and resolve stars to depths $\sim$3 mag fainter than this limit. We do not reach the red giant branch tip. These luminous AGB stars may be the progeny of an intermediate-age population in M32 with lifetimes between 0.2--5 Gyr. 

Using mid-IR photometric criteria, we identify 110 extreme (x-AGB) star candidates, corresponding to approximately half of all sources detected at both 3.6 and 8 $\mu$m. These red stars are highly enshrouded by dust, thus they will often be missed in optical and near-IR surveys. In the Magellanic Clouds the majority of the current dust input into the ISM comes from the extreme AGB stars even though they comprise $<3$ per cent of the population; we expect a similar scenario to occur in M32. Using an assumed gas-to-dust ratio of 200 we estimate the total mass-loss rate from all the x-AGB candidates in M32 to be $1.45 \times 10^{-4}$ ${\rm M}_{\odot} \, {\rm yr}^{-1}$ and measure a cumulative dust-injection rate of  $7.25 \times 10^{-7}$ ${\rm M}_{\odot} \, {\rm yr}^{-1}$.

\section*{Acknowledgements}

The authors wish to thank the referee Greg Sloan for the detailed and relevant comments that have helped improve the clarity of our manuscript.
We would also like to thank Mike Peel, Sundar Srinivasan and everyone involved in the original {\em Spitzer} proposal PID 3400.
OCJ acknowledges the support of an STFC STEP award and thanks the UCLA division of astronomy and ASIAA for their hospitality during the completion of part of this work.
RMR acknowledges support for this work, which was provided by NASA through an award issued by JPL/Caltech to {\em Spitzer} grant GO 3400.
FK acknowledges support from the National Science Council under grant number NSC100-2112-M-001-023-MY3. 
This work is based on observations made with the {\em Spitzer} Space Telescope, which is operated by the Jet Propulsion Laboratory, California Institute of Technology under NASA contract 1407.



\def\aj{AJ}					
\def\actaa{Acta Astron.}                        
\def\araa{ARA\&A}				
\def\apj{ApJ}					
\def\apjl{ApJL}					
\def\apjs{ApJS}					
\def\ao{Appl.~Opt.}				
\def\apss{Ap\&SS}				
\def\aap{A\&A}					
\def\aapr{A\&A~Rev.}				
\def\aaps{A\&AS}				
\def\azh{AZh}					
\def\baas{BAAS}					
\def\jrasc{JRASC}				
\def\memras{MmRAS}				
\def\mnras{MNRAS}				
\def\pra{Phys.~Rev.~A}				
\def\prb{Phys.~Rev.~B}				
\def\prc{Phys.~Rev.~C}				
\def\prd{Phys.~Rev.~D}				
\def\pre{Phys.~Rev.~E}				
\def\prl{Phys.~Rev.~Lett.}			
\def\pasp{PASP}					
\def\pasj{PASJ}					
\def\qjras{QJRAS}				
\def\skytel{S\&T}				
\def\solphys{Sol.~Phys.}			
\def\sovast{Soviet~Ast.}			
\def\ssr{Space~Sci.~Rev.}			
\def\zap{ZAp}					
\def\nat{Nature}				
\def\iaucirc{IAU~Circ.}				
\def\aplett{Astrophys.~Lett.}			
\def\apspr{Astrophys.~Space~Phys.~Res.}		
\def\bain{Bull.~Astron.~Inst.~Netherlands}	
\def\fcp{Fund.~Cosmic~Phys.}			
\def\gca{Geochim.~Cosmochim.~Acta}		
\def\grl{Geophys.~Res.~Lett.}			
\def\jcp{J.~Chem.~Phys.}			
\def\jgr{J.~Geophys.~Res.}			
\def\jqsrt{J.~Quant.~Spec.~Radiat.~Transf.}	
\def\memsai{Mem.~Soc.~Astron.~Italiana}		
\def\nphysa{Nucl.~Phys.~A}			
\def\physrep{Phys.~Rep.}			
\def\physscr{Phys.~Scr}				
\def\planss{Planet.~Space~Sci.}			
\def\procspie{Proc.~SPIE}			
\let\astap=\aap
\let\apjlett=\apjl
\let\apjsupp=\apjs
\let\applopt=\ao


\bibliographystyle{aa}
\bibliography{libby}



 \end{document}